\documentclass[%
 aip,
 jcp,%
 amsmath,amssymb,
 reprint,%
]{revtex4-1}

\usepackage{graphicx}
\usepackage{dcolumn}
\usepackage{bm}
\usepackage{subfigure}
\usepackage [autostyle, english = american]{csquotes}
\MakeOuterQuote{"}
\usepackage{amsmath}

\begin{document}

\preprint{draft-4}

\title{Rotational Dynamics and Angular Locking of Nanoparticles at liquid Interfaces}

\author{Sepideh Razavi}
\affiliation{Department of Chemical Engineering, City College of City University of New York, New York, NY 10031, USA.}%

\author{Ilona Kretzschmar}
 \affiliation{Department of Chemical Engineering, City College of City University of New York, New York, NY 10031, USA.}%

\author{Joel Koplik}
\affiliation{Department of Physics and The Benjamin Levich Institute for Physico-chemical Hydrodynamics, City College of City University of New York, New York, NY 10031, USA.}%

\author{Carlos E. Colosqui}
 \email{carlos.colosqui@stonybrook.edu}
\affiliation{Department of Mechanical Engineering, Stony Brook University, Stony Brook, NY 11790, USA.}

\date{\today}
            
																																%
																																%
																																%
																																%

\begin{abstract}
Nanoparticles with different surface morphologies that straddle the interface between two immiscible liquids are studied via molecular dynamics simulations. 
The methodology employed allows us to compute the interfacial free energy at different angular orientations of the nanoparticle.
Due to their atomistic nature, the studied nanoparticles present both microscale and macroscale geometrical features and cannot be accurately modeled as a perfectly smooth body (e.g., spheres, cylinders).
Under certain physical conditions, microscale features can produce free energy barriers that are much larger than the thermal energy of the surrounding media. 
The presence of these energy barriers can effectively "lock" the particle at specific angular orientations with respect to the liquid-liquid interface. 
This work provides new insights on the rotational dynamics of Brownian particles at liquid interfaces and suggests possible strategies to exploit the effects of microscale features with given geometric characteristics.
\end{abstract}

\maketitle
																																%
																																%
																																%
																																%

\section{\label{sec:Introduction}Introduction}
The dynamic behavior of colloidal particles at fluid interfaces is relevant to a vast field of applications ranging from drug delivery to synthesis of nanostructured materials\cite{Bresme}.
Owing to recent developments in the synthesis of nanoparticles \cite{Glotzer, Xia}, complex geometries and heterogeneous surfaces (e.g., patchy particles) can be engineered to fully exploit physical phenomena such as surface activity \cite{Ramsden,Pickering} and colloidal self-assembly  \cite{Grzybowski,Whitesides}.     
Theoretical models based on continuum thermodynamics are well developed for idealized particle geometries \cite{Pieranski,Binks} such as spheres, ellipsoids, or cylinders.
Recent studies, however, point out significant limitations of such models in the presence of microscale features that are much smaller than the particle and are not considered by idealized geometric representations of the macroscale morphology \cite{Kaz,Carlos}.
These microscale features of physical or chemical nature can cause metastability \cite{Carlos}, which can lead to a very slow relaxation to equilibrium, or even the jamming of single particles at a non-equilibrium state.  
Analytical models considering the transitions between metastable states \cite{Carlos} predict that colloidal particles at fluid interfaces can remain in a given (non-equilibrium) position and angular orientation over unexpectedly long times; this can be highly desirable or undesirable depending on the specific application.
A better understanding of dynamic effects produced by microscale features and surface heterogeneities could enable novel technical applications in which it is desirable to prescribe the translational and rotational motion of the particle.

In the present study, we employ Molecular Dynamics (MD) techniques to investigate the effect of localized features, smaller than the particle size, on the rotational dynamics of nanoparticles that straddle a liquid-liquid interface. 
The article is organized as follows.
In Sec. \ref{sec:Method}, we describe the particle geometries studied and the methodology employed to compute the system free energy as a function of the angular orientation of the particle.
In Sec. \ref{sec:results}, we report main findings and discuss the effects of localized geometric features on the rotational free energy.
The computed free energies reveal that the atomistic nature of the studied nanoparticles produces nontrivial surface morphologies described by an effective diameter that varies as the particle rotates. 
MD simulations show the presence of long-lived metastable states where the particle remains ``locked'' at angular orientations where certain localized features lie at the interface.
The observed lifetime of these metastable states compares well with predictions from Kramers' escape theory \cite{Kramers, Hanggi}.
In Sec. \ref{sec:CR}, we discuss the relevance of our findings for diverse technical applications (e.g., nanoparticle synthesis and self-assembly, reactive emulsions) and possible strategies to exploit the studied effects.

																																%
																																%
																																%
																																%
\section{\label{sec:Method}methodology}
																																%
																																%
																																%
																																%

\begin{figure} [!htbp]
\includegraphics[trim=0cm 0cm 0cm 0cm, clip=true, totalheight=0.23\textheight]{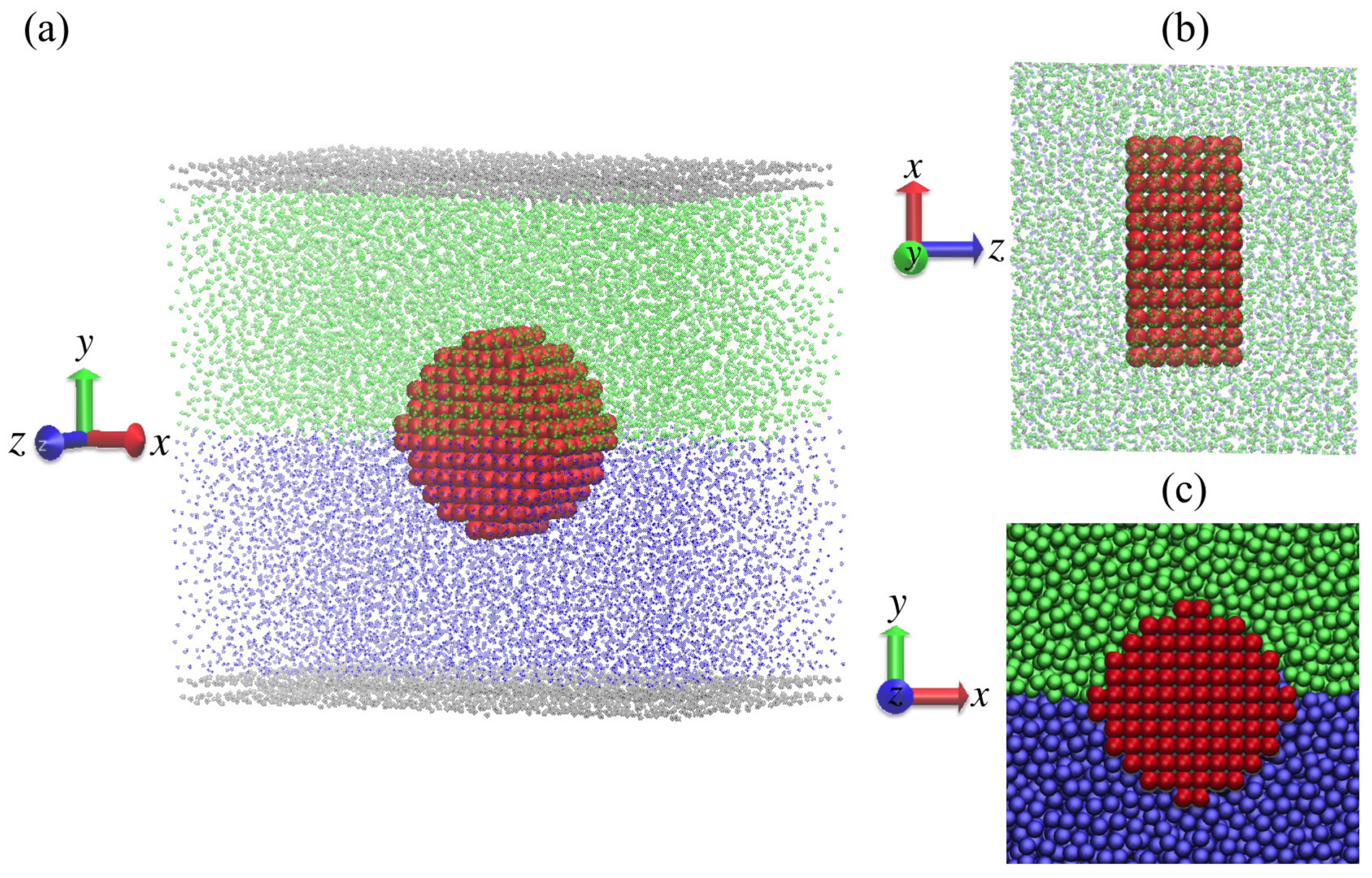}
  \caption{(color online) (a) Snapshot of the simulation box; (b) Top view; (c) Side view. \label{fig:FIGURE1}}
\end{figure}

																																%
																																%
																																%
																																%
\subsection{\label{sec:MD}Molecular Dynamics Simulations}
Numerical simulations in this work are performed via standard MD techniques \cite{Frenkel, Koplik1, Koplik2} employing a generalized Lennard-Jones (LJ) potential 
\begin{equation}\label{eq:lj}
 U(r_{ij})=4\epsilon \left[ \left(\frac{\sigma}{r_{ij}}\right)^{12}-A_{ss'}\left(\frac{\sigma}{r_{ij}}\right)^{6}\right]
\end{equation}
Here, $r_{ij}=|{\bf r_i}-{\bf r_j}|$ is the separation between any two atoms ($i,j=1,N$) of the species $s$ and $s'$, $\sigma$ is the repulsive core diameter, and $\epsilon$ is the depth of potential minimum at $r_{ij}=(2/A_{ss'})^{1/6}\sigma$. 
Following standard procedures \cite{Koplik1, Koplik2} the potential is cut off at $r_c=2.5\sigma$ in order to improve the computational efficiency.
The symmetric coefficient matrix $A_{ss'}=A_{s's}$ determines the intensity of attractive interactions between atoms of different species, which in turn determines the wettability of the particle\cite{Koplik3}. 
The simulated system has four atomic species; fluid 1 $(s=1)$, fluid 2 $(s=2)$, solid particle $(s=3)$, and the surrounding solid wall $(s=4)$. 
The temperature of the system is maintained constant at $T={\epsilon}/{k_B}$ ($k_B$ is the Boltzmann constant) by a Nose-Hoover thermostat, while the number density is $\rho=0.8/\sigma^3$ for all species.
A proper combination of parameters defines a characteristic time scale $\tau = {\sqrt {{m\sigma^2}/{\epsilon}}}$ where $m$ is the atomic mass; the same atomic mass ($m=1$) is employed for species $s=$~1--3, whereas the wall atoms have a larger mass ($m_w=100m$).

In order to model two macroscopially immiscible fluids, we set $A_{12}=A_{21}=0.5$ for cross-interactions, 
while $A_{11}=A_{22}=1$ for self-interactions. 
An interfacial tension of $\gamma=1.5\epsilon/\sigma^2$ is measured for this set of parameters.
The solid particles have symmetric interactions with both liquids ($A_{13}=A_{23}=0.5$) and thus exhibit neutral wetting (i.e., the equilibrium contact angle is $\pi/2$). 
The two fluids are confined by two solid walls located at $y=\pm L_y/2$ and the interface is centered at $y=0$.
Periodic boundary conditions are applied in the $x$ and $z$ directions (see Fig. \ref{fig:FIGURE1}).
%

																															        %
																															        %

\subsection{\label{sec:NP} Effective Size and Morphology of the Nanoparticles}

In this work, we study cylindrical particles in order to characterize the rotational dynamics about a single axis of rotation (i.e., the longitudinal axis of the cylinder).
Furthermore, cylindrical particles are relevant to various applications as they are shown to be more effective stabilizers than spherical and disk-shaped particles \cite{Bon}. 
The modeled particles are composed of $N_p$ atoms that are carved out of an atomic cubic lattice.
The cubic lattice has a spacing of $\Delta x=\Delta y=\Delta z=\rho^{-1/3}$ and the center of each atom is thus located at 
${\bf x}_{ijk}=(\pm x_i, \pm x_j, \pm x_k)$ where $x_i=(i-0.5)\Delta x$; $i,j,k > 0$ are any (nonzero) integers.
The atoms of the three particles are chosen to be within a cylinder of baseline axial length $L_o$~=~6$\sigma$ and three different radii $R$~=~3$\sigma$, 6$\sigma$, and 8.3$\sigma$. 
The simulation domain (cf. Fig.~\ref{fig:FIGURE1}) is a box of dimensions $L_x=6R$, $L_y=20\Delta x+2R$, and $L_z=3L_o$ that scale with the particle radius.

\begin{figure*} [!htbp]
\includegraphics[trim=0cm 0cm 0cm 0cm, clip=true, totalheight=0.57\textheight]{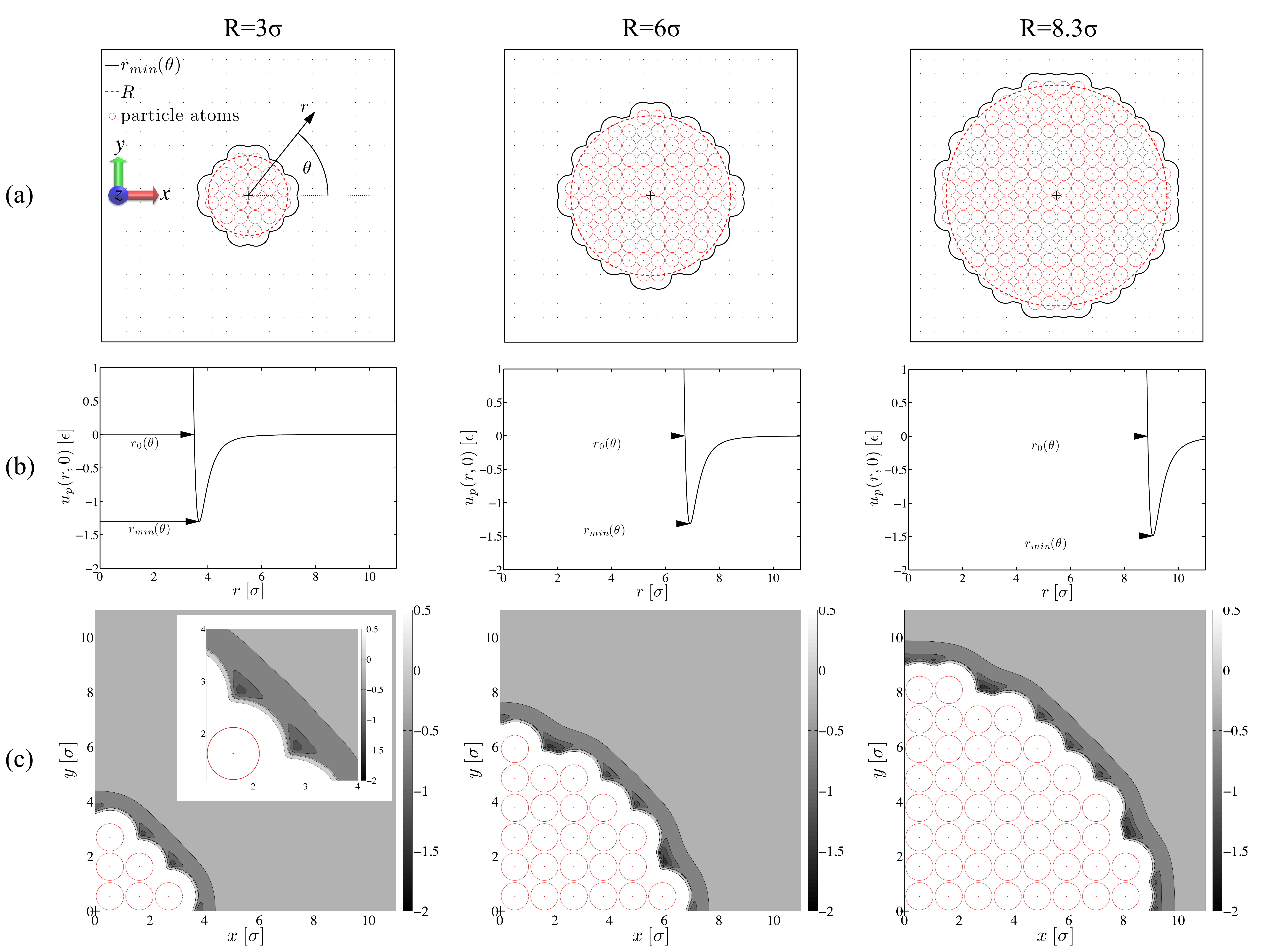}
\caption{(color online) (a) Three studied particles of different size; the particle atoms (red circles) form a cubic lattice and lie within a circle of baseline radius $R$ (dashed red line). 
The curve  $r_{min}(\theta)$ (solid black line) indicates the location of the local minimum of the effective potential $u_p(r,\theta)$ at the particle center ($z=0$). 
The ${\texttt{+}}$ symbol indicates the particle center of mass. 
(b) The effective potential $u_p(r,\theta)$ for $\theta=0$ showing the presence of a local energy well; $u_p(r_{0},\theta)=0$ and $u_p(r_{min},\theta)<0$. 
(c) Two-dimensional contour plots for the effective potential $u_p(r,\theta)$ at the particle center ($z=0$). 
The inset in the leftmost panel is a close-up showing two regions of low energy where $u_p(r,\theta)<0$. 
From left to right: the baseline radius is $R$~=~$3\sigma$, $6\sigma$, and $8.3\sigma$.  
\label{fig:FIGURE2}}
\end{figure*}

The set of baseline radii employed produces the three symmetric cross sections reported in Fig.~\ref{fig:FIGURE2}(a) where the outermost atoms are located at 
$(\pm X_{max},\pm x_{1},\pm x_{3})$ and $(\pm x_{1},\pm X_{max},\pm x_{3})$ with $X_{max}=$~2.5$\Delta x$, 5.5$\Delta x$, and 7.5$\Delta x$, respectively.
A different choice of baseline radii produces different cross sections where the outermost atomic positions do not lie near the main particle axes as discussed later in Sec.~\ref{sec:changes}.
%
																																%
																																%
																																%
																																%
%
																																%
																																%
																																%
																																%

\begin{figure*} [!htbp]
\includegraphics[trim=0cm 0cm 0cm 0cm, clip=true, totalheight=0.43\textheight]{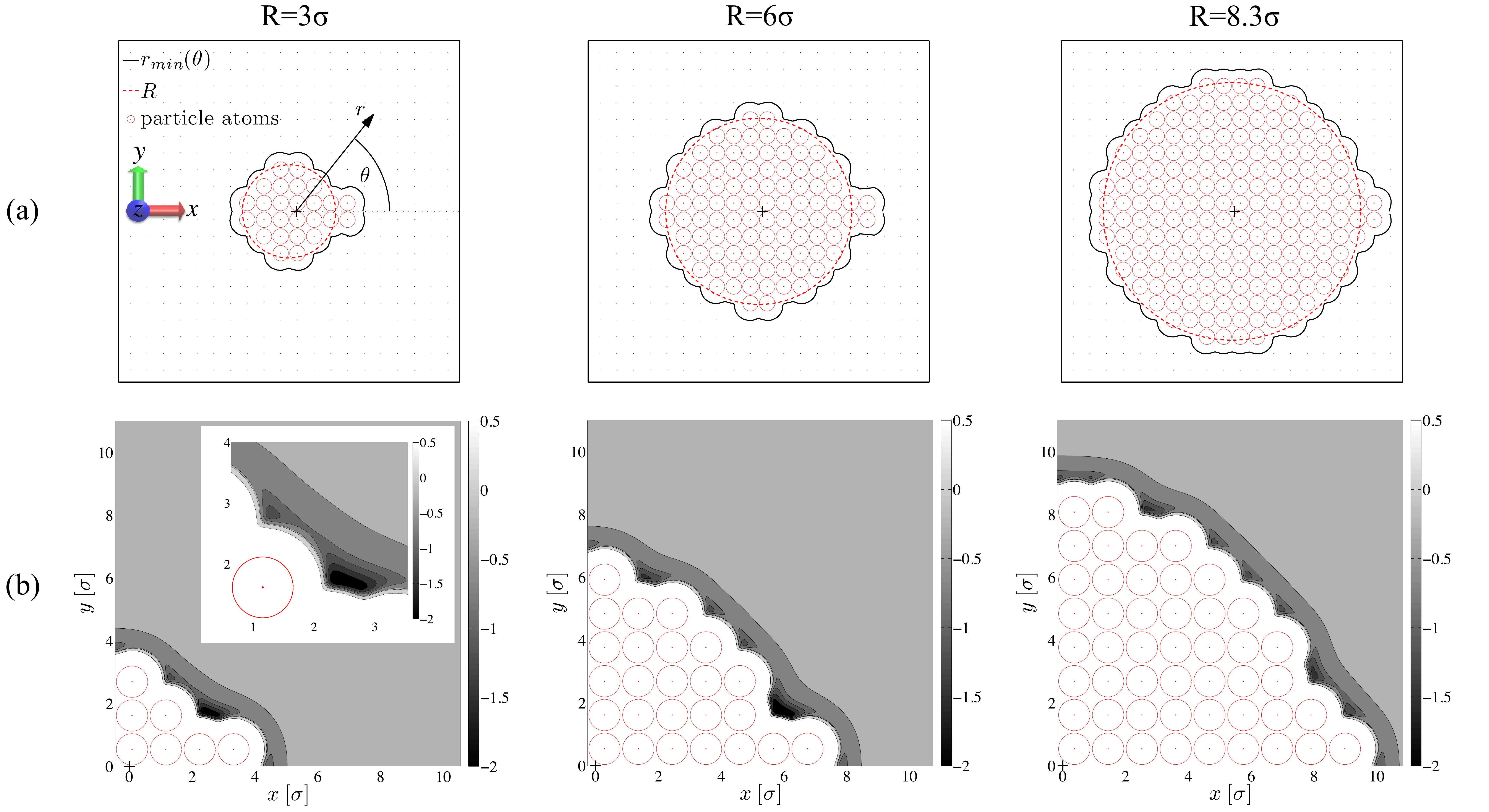}
  \caption{(color online) (a) Three studied particles of different size with a satellite feature, the particle atoms (red circles) form a cubic lattice and lie within a circle of baseline radius $R$ (dashed red line) while a ridge of size $\Delta D\sim1.08\sigma$ is attached. 
The curve  $r_{min}(\theta)$ (solid black line) indicates the location of the local minimum of the effective potential $u_p(r,\theta)$ at the particle center ($z=0$). 
The ${\texttt{+}}$ symbol indicates the particle center of mass. 
(b) Two-dimensional contour plots for the effective potential $u_p(r,\theta)$ at the particle center ($z=0$). 
The inset in the leftmost panel is a close-up showing a large region with low energy $u_p(r,\theta)<0$. 
From left to right: the baseline radius is $R$~=~$3\sigma$, $6\sigma$, and $8.3\sigma$.
\label{fig:FIGURE3}}
\end{figure*}


As described above, the local minimum for the LJ potential employed (Eq.~\ref{eq:lj}) is observed when the interatomic distance is $\sigma_a=(2/A_{13})^{1/6}\sigma\simeq 1.26 \sigma$; this value can be considered as an effective atomic diameter (i.e., the Van der Waals diameter). 
Since the atomic diameter is not negligible when compared to the particle dimensions, we adopt 
$D = 2(X_{max}+\sigma_a)$ as the reference particle diameter, which gives $D\simeq$~$8\sigma$, $14.5\sigma$, and $19\sigma$ for the three studied particles; similarly the effective axial length is $L=2(x_3+\sigma_a)=7.9\sigma$ in all cases.
Since typical values of $\sigma$ range from 0.1 to 0.4 nanometers, the reference diameters of the studied particles correspond to 1 to 8 nanometers in physical units.
To study the effects of localized features, we attach to the three studied particles a satellite feature or ``ridge'' on the surface, consisting of two rows of atoms with length $L$, that protrudes one lattice unit $\Delta x\simeq 1.08\sigma$ from the original surface along the positive $x$-axis as shown in Fig. \ref{fig:FIGURE3}(a). 
The atomistic nature of the studied nanoparticles produces a diversity of localized surface features and the particles cannot be accurately represented by a perfectly smooth geometry (e.g., cylinder).
In order to elucidate the particle morphology it is useful to analyze the effective potential
\begin{equation}\label{eq:u_p}
u_p(r,\theta)=\sum_{j=1}^{N_p} U(|{\bf r}-{\bf r}_j|)
\end{equation}
where ${\bf r}_j$ is the position of each atom in the particle, $r$ is the distance from the center of mass of the particle, and the angular coordinate $\theta$ is here measured in a barycentric system fixed to the particle (see Fig.~\ref{fig:FIGURE2}(a)).
At the axial center of the particle ($z=0$), the surface contour can be approximately represented by the curve $r=r_{min}(\theta)$ along which $\partial u_p/\partial r=0$, since for $|{\bf r}|<r_{min}$ there is a sharp energy increase (see Fig.~\ref{fig:FIGURE2}(b)) and fluid atoms will experience a strong repulsion force. 
Hence the curve $r_{min}(\theta)$ can be used to define a particle diameter
\begin{equation}\label{eq:d_approximation}
d_{min}(\theta)=r_{min}(\theta)+r_{min}(\theta+\pi)
\end{equation}
as a function of the angular orientation $\theta$;
this diameter and the effective particle potential $u_p$ are reported in Fig.~\ref{fig:FIGURE2} and Fig.~\ref{fig:FIGURE3} for the three nanoparticles studied without and with the attached ridge.
%

Besides the irregular nature of the particle surface, the effective interaction potential $u_p$ (cf. Fig.~\ref{fig:FIGURE2}(c) and Fig.~\ref{fig:FIGURE3}(c)) reveals the presence of areas with very low energy $u_p/\epsilon<-1$ to which the fluid atoms will be attracted and potentially remain trapped within a deep energy well.
Below we will discuss how contour irregularities and surface features produce nontrivial effects on the rotational dynamics of the studied nanoparticles.

																																%
																																%
																																%
																																%

\subsection{\label{sec:FEC} Free Energy Computation}
In order to characterize the rotational dynamics of the studied nanoparticles, we compute the Helmholtz free energy as a function of the angular orientation of the particle.
A one-dimensional manifold of the free energy landscape is computed using constrained molecular dynamics \cite{Fernandes,Sepideh}. 
In this type of MD simulation the angular orientation $\theta$ with respect to the fluid-fluid interface is adjusted in a sequence of rotation-relaxation steps; all other degrees of freedom are prescribed, the center of mass of the particle is fixed at $y=0$. 
Each step in the sequence consists of a $\Delta \theta=1^\circ$ rotation about the $z$-axis in a time interval $t_r=500\tau$, after which the angular orientation $\theta$ is constrained during an equally long interval of $t_f=500\tau$. 
During this relaxation interval $t_f$, the time-averaged torque $\ {\langle \mathcal T \rangle}_{\theta^\star}$ about the $z$-axis is measured and the sequence of rotation-relaxation steps is performed for $\theta=[0,2\pi]$.

The Helmholtz free energy $\mathcal{F}$ is determined by the work done on the system during the imposed quasi-static rotation of the particle. 
Changes in the free energy of the system are thus computed by integrating the time-averaged torque:  
\begin{equation}\label{eq:Free}
\Delta\mathcal {F}{(\theta)}=\mathcal{F}(\theta)-\mathcal{F}(\theta_{\rm{ref}})= -\int_{\theta_{\rm{ref}}}^{\theta} \ {\langle \mathcal T \rangle}_{\theta^\star}\: \mathrm{d} \theta^\star
\end{equation}
where $\theta_{\rm{ref}}$ is a reference orientation; for the studied particles it is convenient to choose $\theta_{\rm{ref}}=\pi/2$ as explained below.
%
%
																																%
																																%
																																%
																																%
\begin{figure} [!htbp]
\includegraphics[trim=0cm 0cm 0cm 0cm, clip=true, totalheight=0.43\textheight]{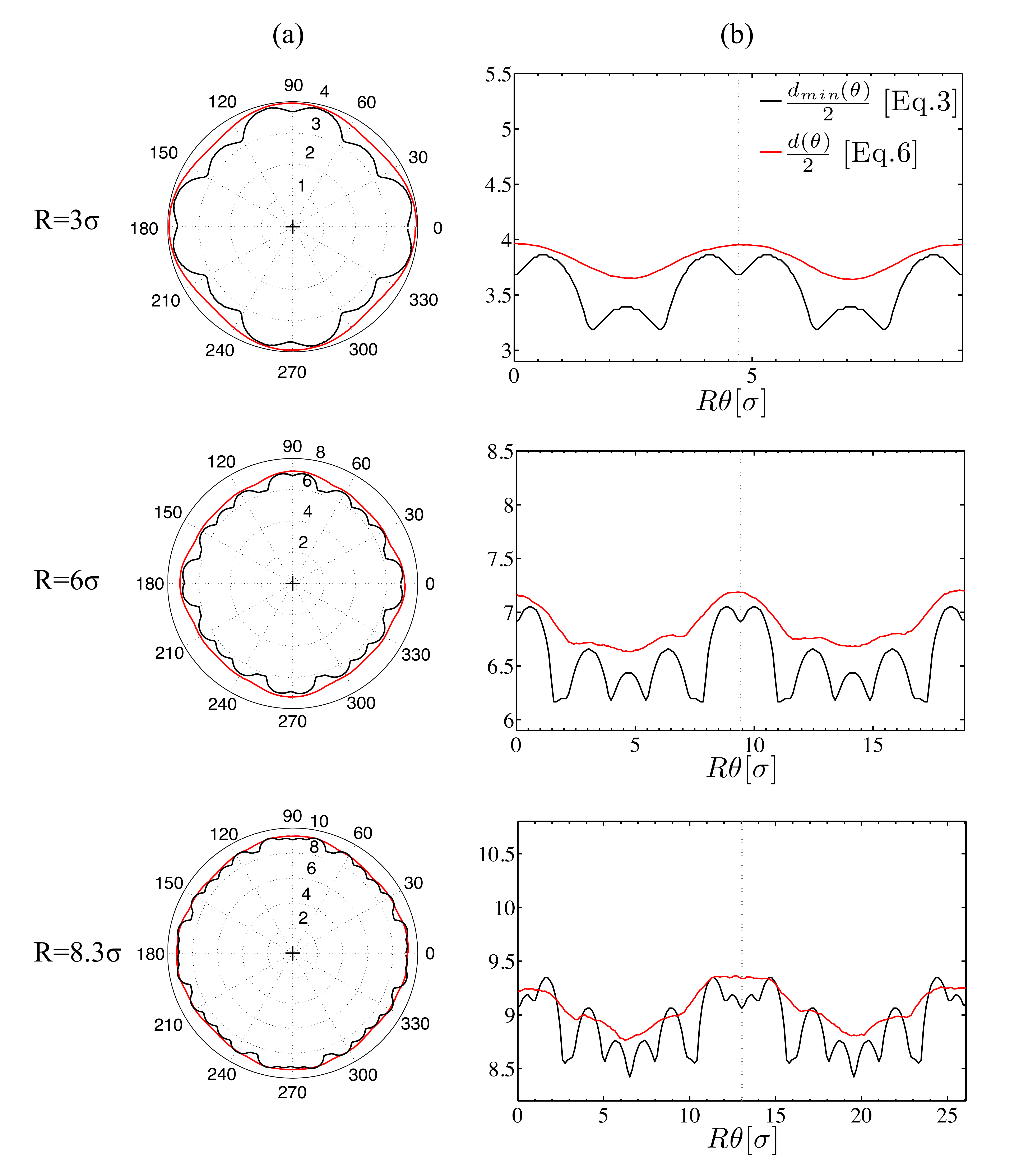}
  \caption{(color online) Particles without a satellite feature: the effective radius $d(\theta)/2$ (red solid line) is computed from MD simulations via Eq.~\ref{eq:d} and $d_{min}(\theta)/2$ (black solid line) is determined from the effective potential $u_p$ at the particle center ($z=0$) via Eq.~\ref{eq:d_approximation}. 
(a) Polar plots showing the effective contours of each particle. 
(b) Linear plots as a function of the arclength $R \theta$ indicating the width of topological features. 
From top to bottom: the baseline radius is $R$~=~$3\sigma$, $6\sigma$, and $8.3\sigma$.
\label{fig:FIGURE4}}
\end{figure}
																																%
																																%
																																%
																																%
\begin{figure} [!htbp]
\includegraphics[trim=0cm 0cm 0cm 0cm, clip=true, totalheight=0.43\textheight]{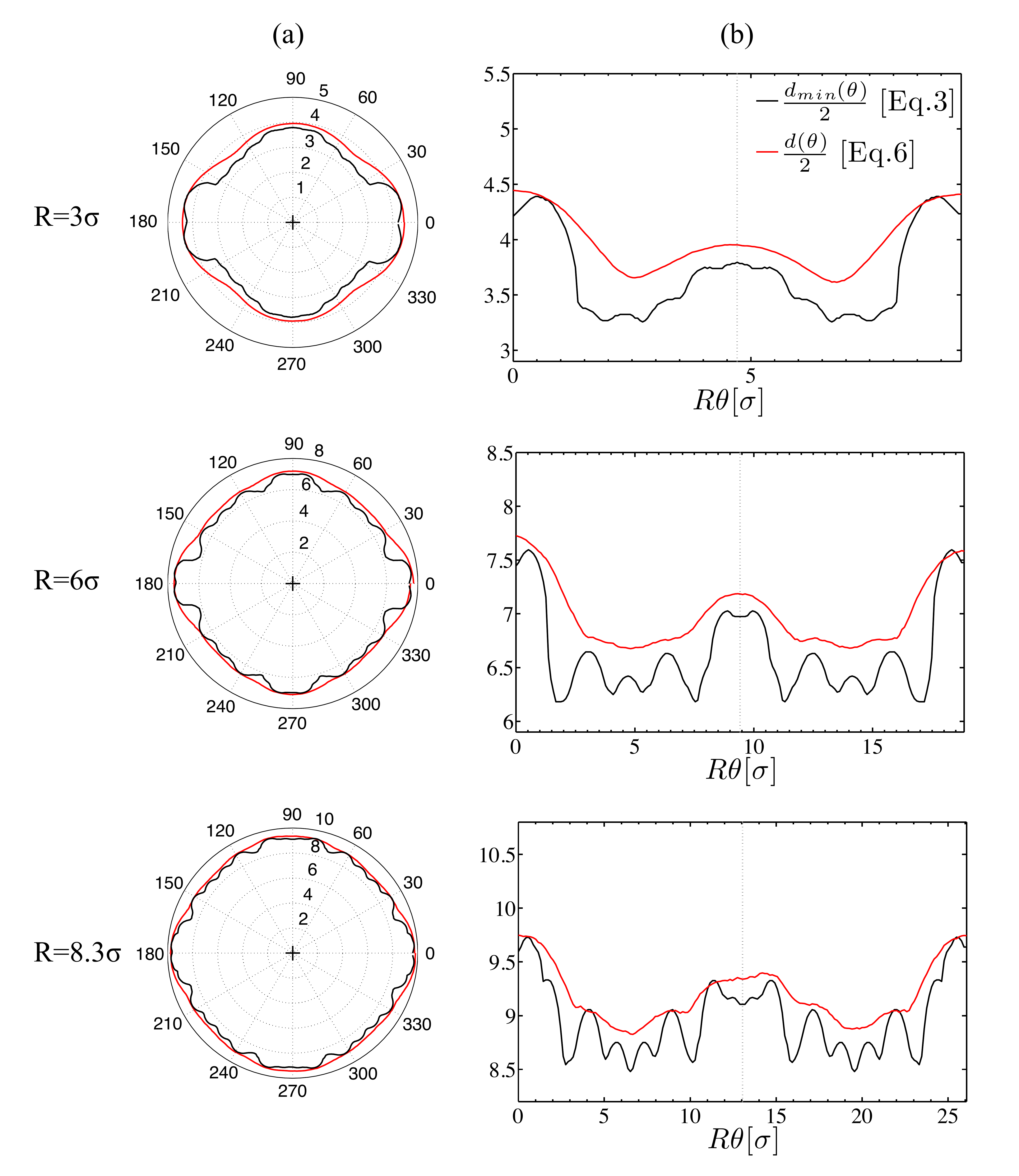}
    \caption{(color online) Particles with a ridge attached at $\theta=0$: the effective radius $d(\theta)/2$ (red solid line) is computed from MD simulations via Eq.~\ref{eq:d} and $d_{min}(\theta)/2$ (black solid line) is determined from the effective potential $u_p$ at the particle center ($z=0$) via Eq.~\ref{eq:d_approximation}. 
(a) Polar plots showing the effective contours of each particle. 
(b) Linear plots as a function of the arclength $R \theta$ indicating the width of topological features. 
From top to bottom: the baseline radius is $R$~=~$3\sigma$, $6\sigma$, and $8.3\sigma$.
\label{fig:FIGURE5}
}
\end{figure}
																																%
																																%
																																%
																																%
\subsection{\label{sec:radius}Free Energy and Effective Diameter}
According to continuum thermodynamics and under the studied conditions (i.e., constant volume and temperature, neutral wetting), changes in the Helmholtz free energy are determined by the area removed from the fluid-fluid interface (i.e., change in interfacial free energy). 
When the particle straddles the interface at a given angle $\theta$, two contact lines are formed at distances $r(\theta)$ and $r(\theta+\pi)$ from the particle center of mass. 
The interfacial area $A(\theta)=L d(\theta)$ removed as the particle rotates is determined by the effective particle diameter $d(\theta)=r(\theta)+r(\theta+\pi)$ given by the distance between the two contact lines.
Assuming a planar interface lying on the horizontal $xz$-plane where $y=0$, the rotational free energy of the studied particles can be cast as
\begin{equation}\label{eq:Helmholtz}
\mathcal{F(\theta)}=C-\gamma L d(\theta)
\end{equation}
where $C$ is an arbitrary constant and $\gamma$ is the interfacial tension of the two liquids. 

Given the symmetry with respect to the $xz$-plane of the three modeled particles, in both cases with or without a satellite feature, we have $d(\pi/2)=2r(\pi/2)=2r(-\pi/2)$. 
Considering the finite size of particle atoms and their interaction range, we employ the approximation $d(\pi/2)= D = 2(X_{max}+\sigma_a)$, where $D$ is the  reference particle diameter defined in Sec.~\ref{sec:NP}, in order to determine an effective particle diameter
\begin{eqnarray}\label{eq:d}
d{(\theta)}= D-\frac{\Delta \mathcal{F}(\theta)}{\gamma L}
\end{eqnarray}
where the free energy difference
\begin{eqnarray}\label{eq:Delta}
\Delta \mathcal{F(\theta)}&=&\mathcal{F(\theta)}-\mathcal{F}(\pi/2) 
\end{eqnarray}
is computed from MD simulations via Eq.~\ref{eq:Free}.  
%
																																%
																																%
																																%
																																%
\section{\label{sec:results}Results and Discussion}
We study three nanoparticles, shown in Fig.~\ref{fig:FIGURE2}, having different effective diameters $D$~=~$8\sigma$, $14.5\sigma$, and $19\sigma$ to which we attach a satellite feature that locally extends the particle diameter by one lattice unit $\Delta D=\Delta x \simeq 1.08\sigma$ as depicted in Fig.~\ref{fig:FIGURE3}(a).
We perform two types of MD simulations; in one approach the particle rotation is constrained to compute the free energy landscape (Sec.~\ref{sec:FEC}), in the other type of simulation the particles are allowed to rotate freely about the $z$-axis (while all other degrees of freedom remain prescribed) to locate different metastable states and measure their lifetimes.

																																%
																																%
																																%
																																%
\begin{figure} [!htbp]
\includegraphics[trim=0.5cm 0cm 0cm 0cm, clip=true, totalheight=0.55\textheight]{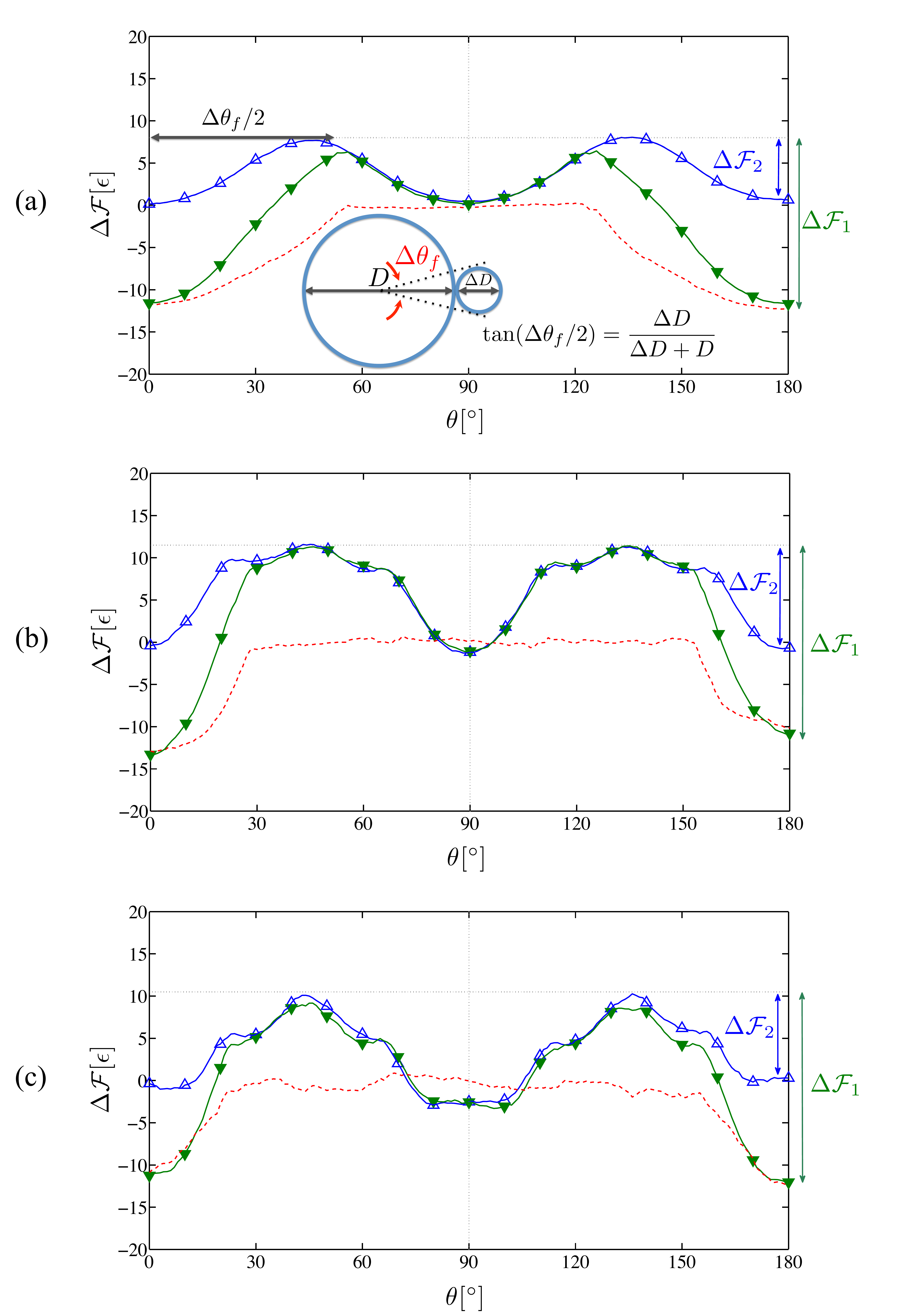}
    \caption{(color online) Free energy landscape of the particles without satellite feature ($\vartriangle$ blue line) and with a satellite feature of size $\Delta D=1.08\sigma$ ($\blacktriangledown$ green line). 
Subtraction of the free energy computed for each case (dashed red line) shows the net effect of the added satellite feature at $\theta=0$. 
From top to bottom (a)-(c): the particle baseline radius is $R$~=~$3\sigma$, $6\sigma$, and $8.3\sigma$.
Panel (a) includes an schematic for estimating the width of the potential well.
\label{fig:FIGURE6}}
\end{figure}

																																%
																																%
																																%
																																%
\subsection{Effective Surface Morphology}
Performing constrained MD simulations we compute the rotational free energy (Eq.~\ref{eq:Free}) and the effective particle diameter $d(\theta)$ (Eq.~\ref{eq:d}) describing the surface morphology of the particle.
The effective diameter $d(\theta)$ and the contour $d_{min}(\theta)$ (Eq.~\ref{eq:d_approximation}) are reported for the three studied particles in Fig.~\ref{fig:FIGURE4} (no satellite feature) and Fig.~\ref{fig:FIGURE5} (satellite feature attached).
Results are presented in polar plots (left panels), which allows us to observe the effective particle shape, and in linear plots as a function of the contour perimeter (right panels), which makes the size of the surface features explicit (e.g., local concavities and convexities).
By definition the effective diameter (Eq.~\ref{eq:d}) is such that $d(\theta)=d(\theta+\pi)$. 
The added ridge (See Fig.~\ref{fig:FIGURE5}) increases the effective diameter of the studied particles by about one atomic diameter $\sigma$ when $\theta=0$ and $\theta=\pi$. 
These two angular orientations correspond to the condition where the center of the satellite feature aligns with the interface between the liquids.  
As it can be observed in Figs.~\ref{fig:FIGURE4}--\ref{fig:FIGURE5}, the effective diameter $d(\theta)$ computed from MD simulations exhibits a smaller number of local minima and maxima than the contour $d_{min}(\theta)$ of minimal potential energy, $u_p$, for single atoms interacting with the particle.
In particular, we note that the minimum distance between neighboring minima and maxima in the profile of $d(\theta)$ is approximately two atomic diameters, $2\sigma$, while the same separation distance for $d_{min}(\theta)$ is approximately one atomic diameter, which corresponds to a surface morphology with local concavities and convexities of smaller size ($\sim\sigma$).
We attribute these observations to multiple fluid-fluid interactions that are not considered in determining the effective particle potential $u_p$ and the contour $d_{min}(\theta)$.
Such fluid-fluid interactions reduce the tortuosity of the effective surface contour $d(\theta)$ through diverse mechanisms such as exclusion volume due to a finite atomic size $\sigma$, and minimization of interfacial energies; e.g., the less tortuous contour $d(\theta)$ requires the production of less interfacial energy as the particle rotates.
Different physical properties of the media correspond to different properties of the fluid-fluid interactions, we thus find that for a nanoparticle the effective diameter to be employed in a continuum-based model for the free energy (Eq.~\ref{eq:Helmholtz}) will depend not only on geometric features but also on physical properties of the media such as surface tension, temperature, mass density, and viscosity. 
%

																																%
																																%
																																%
																																%
  \subsection{\label{subsec:features} Free Energy and Metastability}
In this section, we analyze the rotational free energy obtained from constrained MD simulations via Eq.~\ref{eq:Free} for the three particles described in Sec.~\ref{sec:NP}. 
The values reported in Fig.~\ref{fig:FIGURE6} correspond to the average $[\mathcal F(\theta)+\mathcal F(\theta+\pi)]/2$ measured over a full quasi-static rotation $\theta$~=~$[-\pi,\pi]$. Averaging removes unphysical asymmetries in the rotational energy manifold, attributed to non-equilibrium effects that arise in some of the studied cases (See Appendix).   
For the three modeled nanoparticles (having baseline diameters $3\sigma$, $6\sigma$, and $8.3\sigma$) we compare two study cases: (i) particles without the surface ridge ($\vartriangle$ symbol) and (ii) particles with the surface ridge ($\blacktriangledown$ symbol) that locally extends the particle diameter.
Owing to the symmetric shape of the studied particles the free energies exhibit two pairs of local minima; one pair at $\theta=0$ and $\theta=\pi$, and the other pair at $\theta=\pi/2$ and $\theta=-\pi/2$.
For the particles without the satellite feature or surface ridge, these four minima are the centers of four energy wells with the same depth $\Delta {\mathcal F} \simeq$ 7--12$\epsilon$ and width $\Delta \theta_f \simeq \pi/4$.
For the particles with a satellite feature, the energy wells centered at $\theta$~=~0 and $\theta=\pi$ where the surface ridge intersects the fluid-fluid interface become significantly deeper.
A continuum-based approach expressed by Eq.~\ref{eq:Helmholtz} determines that the net effect of the satellite feature is to increase the depth of these energy wells by $\Delta {\mathcal F}_{f} = \gamma L \Delta D  = 12.76\epsilon$.  
%
																																%
																																%
																																%
																																%
\begin{figure} [!htbp]
\includegraphics[trim=0.5cm 0cm 0cm 0.5cm, clip=true, totalheight=0.39\textheight]{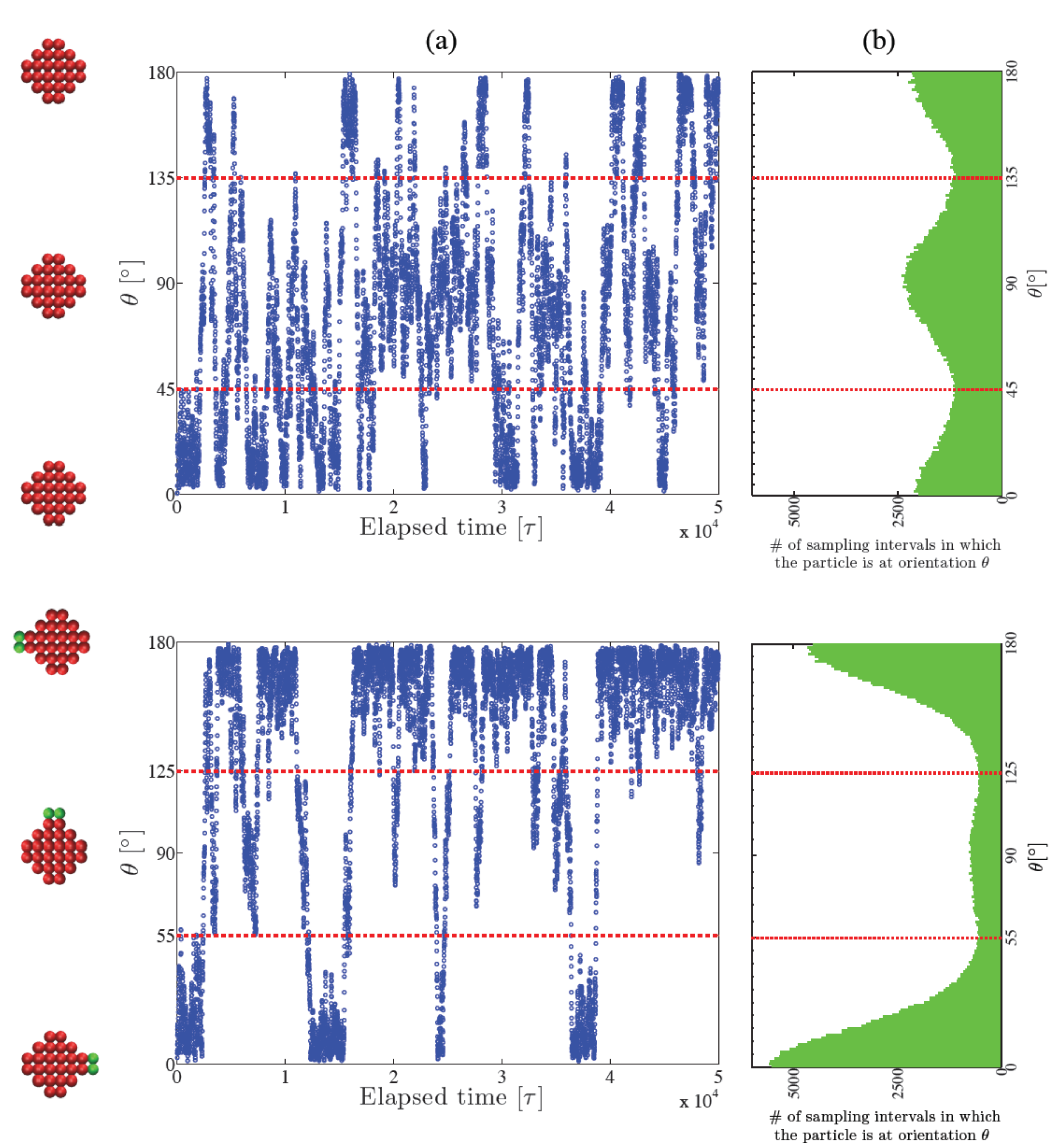}
  \caption{(color online) (a) Simulated particle trajectory $\theta(t)$ for $R=3\sigma$ (three corresponding particle orientations shown on the left). 
(b) Histograms computed using 10 different trajectories showing the probability distribution of various orientations. 
Top panels: particle without added feature. Bottom panels: particle with an attached satellite feature exhibiting metastability.
\label{fig:FIGURE7}}
\end{figure}
Indeed, subtraction of the free energies for cases (i) and (ii) (dotted red curves reported in Fig.~\ref{fig:FIGURE6}) reveals that $\Delta {\mathcal F}_{f} \simeq$~12--13$\epsilon$ which agrees with the prediction via Eq.~\ref{eq:Helmholtz}.
While $\Delta {\mathcal F}_{f}$ is independent of the particle size, the width $\Delta \theta_f$ of the energy wells decreases as the particle diameter increases. 
Using simple geometric arguments (see Fig. ~\ref{fig:FIGURE6}(a)) we find that the well width is approximately 
$\Delta \theta_f\simeq 2\: \mathrm{atan}[1/(1+D/\Delta D)]$; hence, for a fixed feature size the energy wells become narrower as the particle size increases.
 
The local minima in ${\mathcal F}(\theta)$ are separated by maxima located at $\theta\simeq \pi/4$ and $\theta\simeq 3 \pi/4$, which creates well defined energy barriers (cf. Fig.~\ref{fig:FIGURE6}).
When a ridge is attached to the surface, we observe two different energy barriers indicated in Fig.~\ref{fig:FIGURE6} which must be overcome for eight different transitions: $\Delta {\mathcal F}_1\simeq$~20--25$\epsilon$, for the transitions $\theta=0\to\pm\pi/2$ and $\theta=\pi \to \pi\pm\pi/2$; and 
$\Delta {\mathcal F}_2\simeq$~7--14$\epsilon$ for the transitions $\theta=\pi/2\to \pi/2\pm\pi/2$ or $\theta=-\pi/2 \to -\pi/2\pm\pi/2$. 
Thus we find that for the three studied particles with different sizes, the attached satellite feature or surface ridge can produce two degenerate stable states at $\theta=0$ and $\theta= \pi$, where the ridge aligns with the interface, that are longer lived than the two metastable states corresponding to $\theta=\pm \pi/2$, where the ridge does not intersect the interface.  

																																%
																																%
																																%
																																%
\begin{figure} [!htbp]
\includegraphics[trim=0cm 0cm 0cm 0cm, clip=true, totalheight=0.33\textheight]{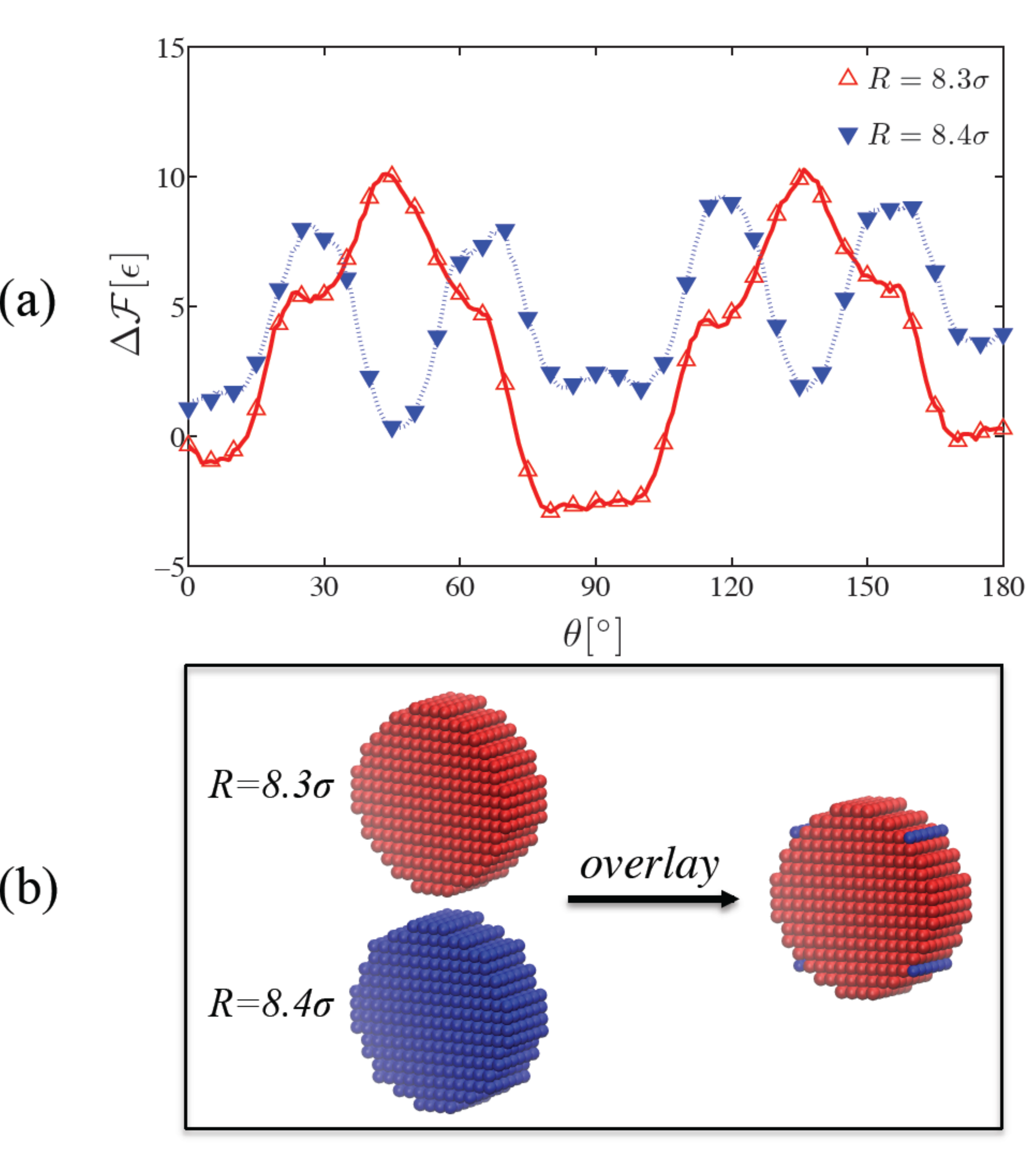}
  \caption{(color online) (a) The free energy landscape for a particle with baseline radius $R=8.3\sigma$ (red $\vartriangle$) and $R=8.4\sigma$ (blue $\blacktriangledown$). (b) Difference in particle atomic structure going from $R=8.3\sigma$ to $R=8.4\sigma$ leading to the additional local minima.\label{fig:FIGURE8}}
\end{figure}
																																%
																																%
																																%
																																%
\subsection{Escape From a Metastable State}
The rotational free energy of the studied nanoparticles exhibits significant energy barriers that can produce long-lived metastable states when the particle undergoes purely rotational Brownian motion. 
In order to study these metastable states we perform standard MD simulations at three different temperatures $T=$~2, 2.5, and 3$\epsilon/k_B$ where the angular orientation $\theta$ evolves freely in time.
The trajectory $\theta(t)$ for the nanoparticle with effective diameter $D=8\sigma$ is shown for the cases without and with the satellite feature in Fig.~\ref{fig:FIGURE7}(a) and ~\ref{fig:FIGURE7}(b), respectively.
As predicted in the analysis of the computed free energy landscapes via constrained MD simulations, the satellite feature produces two stable states at $\theta=0$ and $\theta=\pi$ with equal average lifetime $\Lambda_1$.
Two additional metastable states with shorter lifetime $\Lambda_2<\Lambda_1$ are observed for $\theta=\pm \pi/2$ as can be seen from the histograms for the trajectory $\theta(t)$ (rightmost panels in Fig.~\ref{fig:FIGURE7}).

																																%
																																%
																																%
																																%
\begin{table*}[!htbp]
\caption{Comparison between the life-time ratios $\Lambda_1/\Lambda_2$ calculated from MD simulations and Kramers' theory (\ref{eq:kramers2}).} 
\centering
\begin{minipage}{14cm}
\begin{tabular}{c|c|c|c|c|c|c|c|c|c}
 \hline\hline

 $T (\epsilon/k_B)$  & $\gamma (\epsilon/\sigma^2)$ & \multicolumn{2}{c} {$R=3\sigma$} & \multicolumn{2}{c} {$R=3\sigma$  (with feature $\Delta D$)}& \multicolumn{2}{c}{$R=6\sigma$} &\multicolumn{2}{c}{$R=6\sigma$ (with feature $\Delta D$)}
 \\ [0.5ex]
\hline
 &  & MD & Theory  & MD & Theory & MD & Theory & MD & Theory
  \\ [0.5ex]
\hline
 3& 0.74 & 1.08 & 1  & 5.52 & 5.56 & 1.1 & 1 & 4.5 & 5.12
  \\ [0.5ex]
   2.5& 1.02 & 1.1& 1  & 14.5& 14.76 & 1.17 & 1 & 20.1 & 19.75
  \\ [0.5ex]
     2.0& 1.28 & 1.4& 1  & 75.4 & 29.5 & 1.17 & 1 &$\times$ & 137
  \\ [0.5ex] 
       1.0& 1.49 & \:\:$\times$ \footnote{$\times$ symbol indicates that the available computation time is not enough for statistically relevant sampling.}& 1  & $\times$& 1.64E${\texttt{+}}$04 & $\times$ & 1 &$\times$ & 6.4E${\texttt{+}}$04
  \\ [0.5ex] 
\hline\hline

\end{tabular}
\end{minipage}
\label{tab:hop}
\end{table*}
%
																																%

The lifetime of these (meta)stable states can be estimated using Kramers' theory for thermally-activated transitions \cite{Kramers, Carlos}:
\begin{equation} \label{eq:kramers}
\Lambda_{k}=C_k \exp\left(\frac{\Delta {\mathcal F}_k}{k_B T} \right)\quad(k=1,2)
\end{equation}
Here ${\Delta \mathcal F}_k$ is the energy barrier and $C_k$ is a prefactor given by Kramers' theory. 
The prefactor $C_k=2\pi\xi /\sqrt{\ddot{{\mathcal F}}_{min}|\ddot{{\mathcal F}}_{max}|}$ in Eq.~\ref{eq:kramers} is determined by the rotational viscous damping $\xi$ of the particle and the local curvature of the free energy $\ddot{{\mathcal F}}\equiv \partial^2  {\mathcal F}/\partial \theta^2$ at the minima and neighboring maxima for each metastable state \cite{Carlos,Hanggi}.
Hence, one can obtain the ratio of the expected lifetimes as
\begin{equation} \label{eq:kramers2}
\frac{\Lambda_1}{\Lambda_2}=\sqrt{\frac{\ddot{{\mathcal F}}(\pi/2)}{\ddot{{\mathcal F}}(0)}} \exp\left(\frac{\Delta {\mathcal F}_1-\Delta {\mathcal F}_2}{k_B T} \right)
\end{equation}
We computed the ratio $\Lambda_1/\Lambda_2$ in MD simulations for the two smallest particles, where $D$~=~$8\sigma$ and $14.5\sigma$, by performing ensemble average over 10 different realizations that run for a simulation time of $10^6 \tau$.
These numerical results are reported in Table 1 and compared against theoretical predictions from Eq.~\ref{eq:kramers2}.
There is a good agreement between numerical results and theoretical predictions for $k_B T/\epsilon$~=~2.5 and 3. 
The agreement deteriorates as the temperature is further reduced because long simulation times, much larger than $\Lambda_1$, that are required to observe multiple transitions and perform accurate statistical sampling are not accessible for $k_B T/\epsilon<2$.
For all studied conditions we observe that particles to which the satellite feature is attached remain over significantly longer times in the angular orientation $\theta=0$ or $\theta=\pi$ where the satellite feature binds to the liquid-liquid interface.
Owing to the exponential factor in Eq.~\ref{eq:kramers2}, the time during which the particle remains at angular orientations $\theta=0\pm\Delta \theta_f$ or $\theta=\pi\pm\Delta \theta_f$ is extremely sensitive to increasing the energy barrier difference $(\Delta {\cal F}_1-\Delta {\cal F}_2)$ and/or decreasing the thermal energy. 
Indeed, MD simulations for $k_B T/\epsilon<2$ performed over a simulation time $10^6\tau$ do not show thermally-activated transitions once the angular trajectory reaches $\theta=0$ or $\theta=\pi$. 
Hence, the angular locking produced by the attached satellite feature can be employed to maintain the particle at specific angular orientations with respect to the interface within a given range of system temperatures.
%

																																%
																																%
																																%
																																%
\subsection{Effect of Changes in the Atomic Structure \label{sec:changes}}
All particles studied in Fig. \ref{fig:FIGURE2} have four local minima of the free energy caused by fluctuations of their effective diameter $d(\theta)$.
The number of minima, however, is sensitive to slight variations of the baseline radius $R$ employed in defining the particle atoms that form a cubic lattice. 
To illustrate this effect, the free energy of two particles with baseline radii of $R=8.3\sigma$ and $R=8.4\sigma$ are  shown in Fig.~\ref{fig:FIGURE8}(a).  
As it can be observed, the particle of radius $R=8.3\sigma$ has four minima (three located at $\theta=0$, $\theta=\pi/2$, $\theta=\pi$ are shown in Fig.~\ref{fig:FIGURE8}(a)) that create energy wells with depth of $\Delta \mathcal F$ $\sim 10\epsilon$ (as discussed in Sec.~\ref{subsec:features}). 
On the other hand, the particle with radius $R=8.4\sigma$ has a different free energy landscape with five local minima as seen in Fig. \ref{fig:FIGURE8}(a).
The different number of local minima can be explained by comparing the shape of the two particles shown in Fig. \ref{fig:FIGURE8}(b). 
Overlaying the two structures evidences that a slight increase in local radius of the particle, originating from the additional atoms, alters the effective local diameter and thus the free energy landscape. 
Therefore, microscale surface features are likely to produce changes in the rotational dynamics of particles at interfaces.
However, if an intentionally added feature has a much larger effect than microscale features or atomic scale defects, the particle will still be locked at a desired angular orientation.

																																%
																																%
																																%
																																%

\section{\label{sec:CR}CONCLUDING REMARKS}
Molecular dynamics simulations of nanoparticles at the interface of two immiscible liquids revealed important dynamical effects associated with the localized geometric features that produce local minima in the system free energy.
These effects can be described by the proposed analytical models, based on continuum thermodynamics when considering microscopic features such as a finite atomic radius and the atomistic surface morphology.  
In particular, we found that the nanoparticle geometry can be designed so that its rotational dynamics present long-lived metastable states at certain angular orientations with respect to a fluid-fluid interface.
The lifetime of these metastable states can be predicted from a detailed knowledge of the particle geometry by using the employed analytical model based on Kramers' theory for thermally-activated transitions.
Our findings suggest that nanoparticles with specific geometries could thus be synthesized for a number of technical applications where it is desirable to tame Brownian effects.

One such application is a particle-stabilized reactive emulsion where reactions take place in a biphasic mixture of two immiscible liquids\cite{Crossley}, while colloidal particles can be tuned for dual functionality (catalysis and stability). 
Emulsion stabilization can be accomplished through particle binding to liquid-liquid interfaces, whereas reactivity can be directed to a specific phase by selective functionalization of the particle surface. 
In order to avoid competing reactions from hindering product formation and to achieve high yield \cite{Lee}, the catalytically active surface of the particle must face the proper liquid phase. 
Reaction yield and specificity would be dramatically increased if active surface groups face the preferred phase, the phase with which they chemically react, over long periods of time.     
Hence, enhancing the emulsion reactivity could be accomplished with nearly spherical or cylindrical nanoparticles having an equatorial ridge, similar to the one modeled in this work, and reactive groups located near the poles.
As shown in our study, such nanoparticles will remain locked near the optimal orientation for enhanced reactivity when the system temperature is reduced below a given threshold value.
Other important applications where similar ideas can be exploited are synthesis of patchy particles, drug delivery, surface catalysis, and interfacial self-assembly.
Future work will include studying the coupling between translational and rotational degrees of freedom (see Appendix), which requires the computation of multidimensional free energy landscapes.  
%

																																%
																																%
																																%
																																%
\begin{acknowledgments}
This work was partially supported by the City University of New York High Performance Computing Center under NSF Grants CNS-0855217 and CNS-0958379.
Support from the National Science Foundation through awards CBET-1067501 and PREM (DMR-0934206) is acknowledged.
\end{acknowledgments}

																																%
																																%
																																%
																																%
\appendix*
\section{}
\renewcommand\thefigure{\thesection\:A.\arabic{figure}}    
\setcounter{figure}{0}

																																%
																																%
																																%
																																%
\begin{figure*} [!htbp]
\includegraphics[trim=0cm 0cm 0cm 0cm, clip=true, totalheight=0.5\textheight]{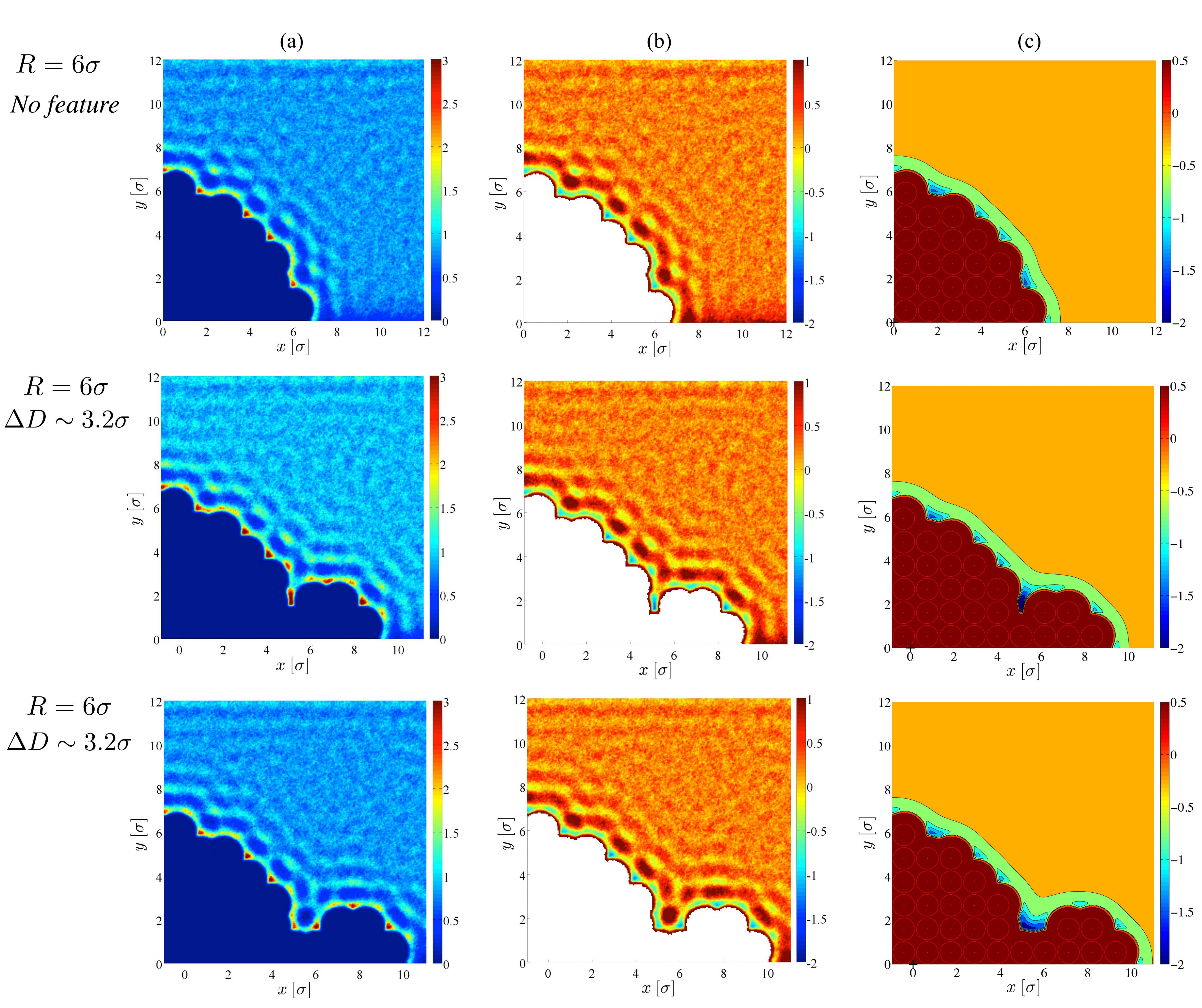}
  \caption{(color online) (a) Number density $\rho$ from MD calculations for $\theta=0$. 
(b) Mean field potential $u_m=-k_BT \log\rho$ from MD calculations for $\theta=0$.
(c) Effective interaction potential $u_p$ between the particle and a single fluid atoms. 
From top to bottom: particle with no satellite feature and particles with satellite feature of $\Delta D\sim 3.2\sigma$ at two different distances from the surface.  \label{fig:FIGUREA1}}
\end{figure*}
																																%
																																%
																																%
																																%
\subsection{\label{density}Density Profiles and Mean Potential Field}
The effective interaction potential between a particle and a single fluid atom was introduced in Sec.~\ref{sec:radius} in order to study nontrivial features of the particle morphology. 
It was stated that fluid atoms could be trapped within the observed regions of low energy where $u_p/k_BT<-1$, which could cause contact line pinning and other effects.  
Here, we compare the effective interaction potential $u_p$ against the number density profile $\rho$ and mean field potential $u_m=-k_B T \log \rho$ computed from MD simulations.
Results are reported in Fig.~\ref{fig:FIGUREA1} for three different particles with the baseline radius $R=6\sigma$ and two satellite features of size $\Delta D=3.2 \sigma$ that are attached at different distances from the particle surface. 
As seen in Fig. \ref{fig:FIGUREA1}, crucial features of the effective potential $u_p$ (e.g., surface irregularities, low energy spots) are preserved in the number density and mean field where multiple fluid-fluid interactions are considered.
From the number density profiles we readily observe that fluid atoms are indeed trapped within regions where the effective particle potential is very low, significantly raising the local number density. 
The main effect not captured by the effective particle potential $u_p$ is a layering of the liquid atoms that leads to density fluctuations in the vicinity of the particle surface. 
These fluctuations, reported also in previous work \cite{Sepideh}, decay farther away from the particle surface.

For cases with the added feature, a large region of low energy appears in the gap between the particle surface and the satellite feature as the ridge is attached farther from the particle surface.
This gap can thus cause pinning of the three-phase contact line at certain angular orientations, which can entail additional ``locking'' effects as discussed below. 
%
%

																																%
																																%
																																%
																																%

\begin{figure*} [!htbp]
\includegraphics[trim=0cm 0cm 0cm 0cm, clip=true,totalheight=0.32\textheight]{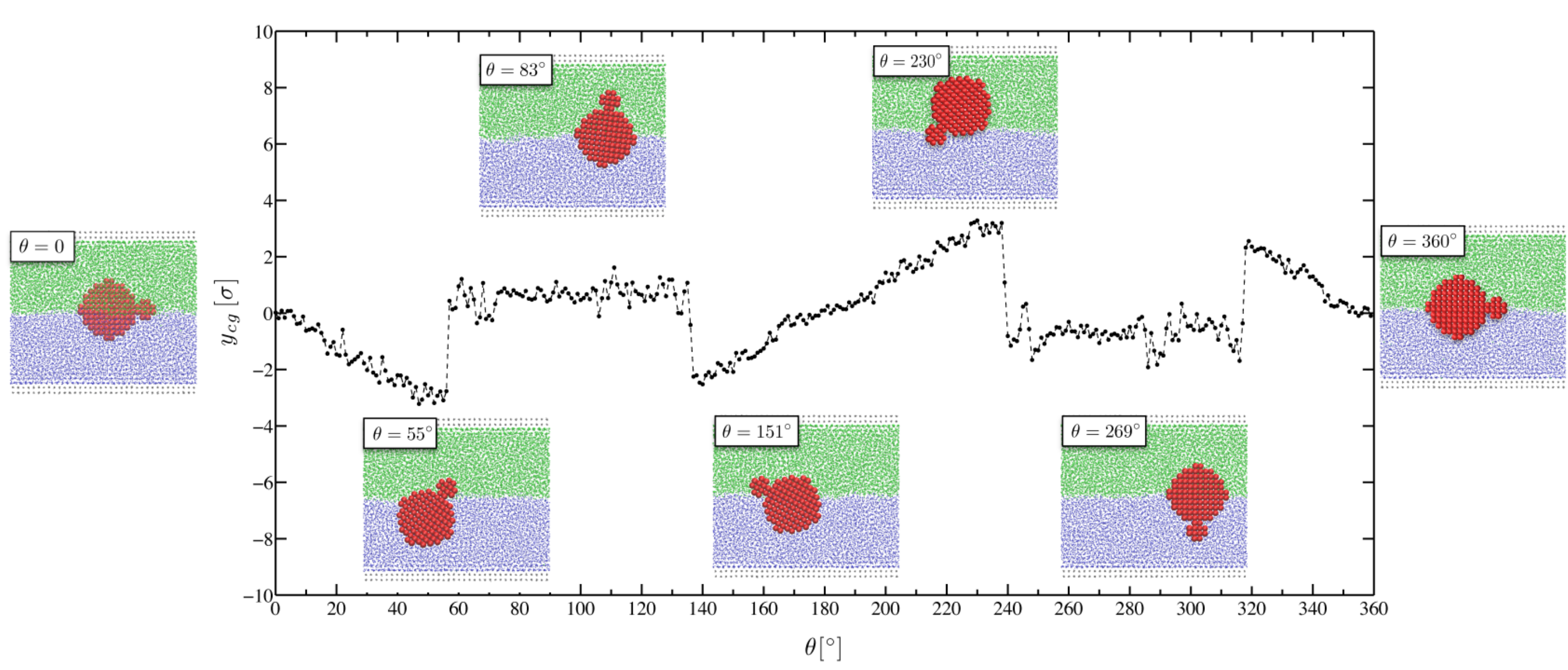}
\caption{(color online) Particle center of mass $(y_{cg})$ measure with respect to the interface $(y=0)$ versus angular orientation $\theta$. 
The baseline radius of the particle is $R=6\sigma$ and the satellite feature extends the local diameter by $\Delta D = 2\Delta x\sim 2.15\sigma$.
Snapshots for different configurations are shown. \label{fig:FIGUREA2}}
\end{figure*}

																																%
																																%
																																%
																																%

\subsection{Rotation-Translation Correlation}
As explained in section \ref{sec:MD}, thermodynamic integration is performed via quasistatic rotation of the particle that straddles the liquid-liquid interface.
In the calculation of the Helmholtz free energy reported in Fig.~\ref{fig:FIGURE6}, the particle is held fixed at the interface center while incremental changes in $\theta$ are imposed. 
When the particle is allowed to translate freely in the direction normal to the interface, a correlation between the translation and rotation of the particle is significant for some particles.
This correlation is illustrated in Fig.~\ref{fig:FIGUREA2} in which the sink (or rise) of the particle occurs at specific angular orientations where the contact line pins at (or de-pins from) the particle-feature gap.
At each angular orientation the particle adjusts its center of mass adopting the configuration that minimizes the free energy of the system. 
This configuration is the one that produces a minimal interfacial energy either by removing fluid-fluid interfacial area or reducing the curvature of the interface.
As the contact line pins at the particle-feature gap for angular orientations where the feature is close to the fluid interface (see Fig.~\ref{fig:FIGUREA2}), the particle sinks or rises to prevent the creation of additional interfacial area.

																																%
																																%
																																%
																																%
\bibliography{aipsamp}
\footnotesize{

\end{document}